\documentclass[]{spie}  

\usepackage{amsmath,amsfonts,amssymb}
\usepackage{graphicx}
\usepackage{subfigure}
\usepackage[colorlinks=true, allcolors=blue]{hyperref}

\title{Introduction to the 7-Dimensional Telescope: Commissiong Procedures and Data Characteristics}

\author[a, b]{Ji Hoon Kim}
\author[a, b]{Myungshin Im}
\author[a, b]{Hyung Mok Lee}
\author[a, b]{Seo-Won Chang}
\author[a, b]{Hyeonho Choi}
\author[a, b]{Gregory S. H. Paek}
\affil[a]{Astronomy Research Center, Seoul National University (SNU), Seoul, Republic of Korea}
\affil[b]{Astronomy Program, Department of Physics and Astronomy, SNU, Seoul, Republic of Korea}

\authorinfo{Further author information: (Send correspondence to Ji Hoon Kim.)\\Ji Hoon Kim: jhkim.astrosnu@gmail.com,
Myungshin Im: myungshin.im@gmail.com}

\begin{document} 
\maketitle

\begin{abstract}
The 7-Dimensional Telescope (7DT) is a multi-telescope system designed to identify electromagnetic (EM) counterparts of gravitational-wave (GW) sources.
Consisting of 20 50-cm telescopes along with 40 medium-band filters of 25 nm width, 7DT can obtain spectral mapping images for a large field of view ($\sim$1.25 square degrees).
Along with flexible operation, real-time data reduction, and analysis, the 7DT's spectral mapping capability enables 7DT to follow up GW events quickly and discover EM counterparts.
Among 20 planned telescopes, 12 units are deployed at the El Sauce Observatory located at Rio Hurtado Valley in Chile.
Since we obtained the first light of 7DT in October 2023, we started its commissioning procedures including examination of bias levels, master flat production, and spectrophotometric standardization.
In this talk, we present 7DT instruments and their set-up, commissioning procedures, and data characteristics of 7DT along with our three-layered surveys which are assumed to be initiated in early 2024.
\end{abstract}


\section{INTRODUCTION}
\label{sec:intro}  

The first discovery of the gravitation-wave (GW) electromagnetic counterpart in 2017 ignited the field of multimessenger astronomy (MMA)\cite{2017ApJ...848L..12A}.
However, it remains the lone case.
Kilonovae (KNe) are the end products of compact binary coalescence events including at least one neutron star (NS) and have been quite rare and elusive.
There are three obstacles that make them difficult to detect; their low luminosities, large localization areas based on GW signals, and numerous transients and spurious objects within localization areas.
Typical KNe have absolute magnitudes around -15 to -17 at their peak, evolving much more rapidly than Type Ia supernovae, (SNe Ia) with timescales of a week or so.
Moreover, their brightness decay fast; roughly 0.5 mag per day in the optical wavelengths.
During the O3 run of the gravitational-wave detector network that consists of the Advanced Laser Interferometer Gravitational-wave Observatory\cite{2015CQGra..32g4001L}, the Advanced Virgo\cite{2015CQGra..32b4001A}, and the Kamioka Gravitational Wave Detector (KAGRA)\cite{2021PTEP.2021eA101A}, typical localization areas range from hundreds to thousands of square degrees\cite{2020LRR....23....3A}.
Having started on May 24, 2023, the O4 run has yet to see much improvement over this, although an order-of-magnitude improvement was expected.
Such large localization areas are not only difficult to cover with optical telescopes but also follow-up observations have a huge number of spurious objects, including other types of transients as well as detector artifacts.
In a FoV of 100 square degrees, the number of transient objects including SN Ia, compact binary coalescences, and classical novae is expected to be around 100 per 7 days\cite{2015ApJ...814...25C}.

To overcome these hurdles, optical follow-up observation facilities should be able to cover lots of areas as soon as possible.
Therefore, these facilities should utilize telescopes with a large FoV and their operation must be flexible. 
On top of these requirements, spurious objects should be vetted to confirm true GW electromagnetic counterparts.
While ruling out detector artifacts is tedious, but trivial, separating KNe from other transients needs spectroscopic classification, albeit with low spectral resolution.


There are several facilities of multi-telescope arrays to address the limited FoV issue for GW event follow-up observations.
These multi-telescope arrays utilize commercial-off-the-shelf optical tube assemblies (OTAs) and mounts which are mass-produced for non-research activities taking advantage of cost efficiency and quick implementation. 

The Gravitational-wave Optical Transient Observer (GOTO) adopts 4 arrays of 8 commeercial off-the-shelf (COTS) optical telescopes located at two sites; one in the Northern Hemisphere and the other in the Southern Hemisphere\cite{2022MNRAS.511.2405S}.
Utilizing eight 40-cm reflectors with a focal ratio of f/2.5, each array can cover about 40 square degrees with an exposure.
Operated by the Monash-Warwick Alliance which includes University of Warwick, Monash University, and some other partner institutes around the World, GOTO uses the four Baader L, G, R, B filters.

Located at the European Southern Observatory (ESO) La Silla site in Chile, BlackGEM currently consists of three 65-cm telescopes\cite{2022SPIE12182E..1VG}.
With a focal ratio of f/5.5, each telescope covers 2.7 square degrees, which makes the total coverage by the entire array 8.1 square degrees.
BlackGEM uses Sloan broad-band filters of u-, g-, r-, i-, and z-bands and an additional wide-band of q-band filter.
BlackGem is designed, built, and operated by the BlackGEM Consortium which was founded by the Netherlands Research School of Astronomy (NOVA), Radboud University, and the KU Leuven.

The Large Array Survey Telescope (LAST) consists of 48 27.9-cm COTS telescopes with a focal ratio of f/2.2\cite{2023PASP..135f5001O}.
Designed, built, and operated by Weizmann Institute of Science, LAST is located at the Weizmann Astrophysical Observatory site in the Israeli Negev desert.
Each telescope covers a FoV of 7.4 square degrees, therefore the entire system of LAST provides a FoV of 355 square degrees when its telescopes are in divergent mode.
LAST carries out filterless follow-up observations and its photometry is calibrated against the GAIA catalog.\cite{2021A&A...649A...1G}


In this work, we introduce the 7-Dimensional Telescope (7DT) that is designed, built, and operated by Center for the Gravitational-wave Universe at Seoul National University.
A multi-telescope array with 20 50-cm reflectors, 7DT provides low-resolution spectroscopy for objects within a large field of view (FoV). 
Combined with its flexible operation capability and real-time data reduction and analysis pipeline, 7DT is capable of capturing candidates of GW electromagnetic counterparts and classifying them.
In the following sections, we describe the overview of observing and computing facilities, data acquisition and reduction software, and commissioning procedures.
We also present prospects including our survey plan.

\section{System Overview}
\label{sec:hardware}

7DT consists of 20 50-cm COTS telescopes with a focal ratio of f/3.
Each telescope is equipped with a CMOS camera whose SONY IMX455 sensor has 9576 by 6388 pixels of 3.76 $\mu m$.
At f/3, the camera provides a field of view of 1.33 by 0.89 degrees with a pixel size of 0.5 by 0.5 arcseconds.
Each telescope is equipped with Sloan g-, r-, and i-band filters while Sloan u-, and z-band filters are available for a few units.
On top of these broad-band filters, 7DT utilizes 40 medium-band filters with a full-width half maximum (FWHM) of 25 nm.
These medium-band filters cover the wavelength range from 400 nm through 900 nm.
Currently, 12 telescopes are deployed at the site in Chile and are operational with 2 sets of 20 medium-band filters whose central wavelengths range from 400 nm to 875 nm with a gap of 25 nm.

\begin{figure} [ht!]

\begin{center}
\begin{tabular}{cc} 
\subfigure[]{\includegraphics[width=0.48\textwidth]{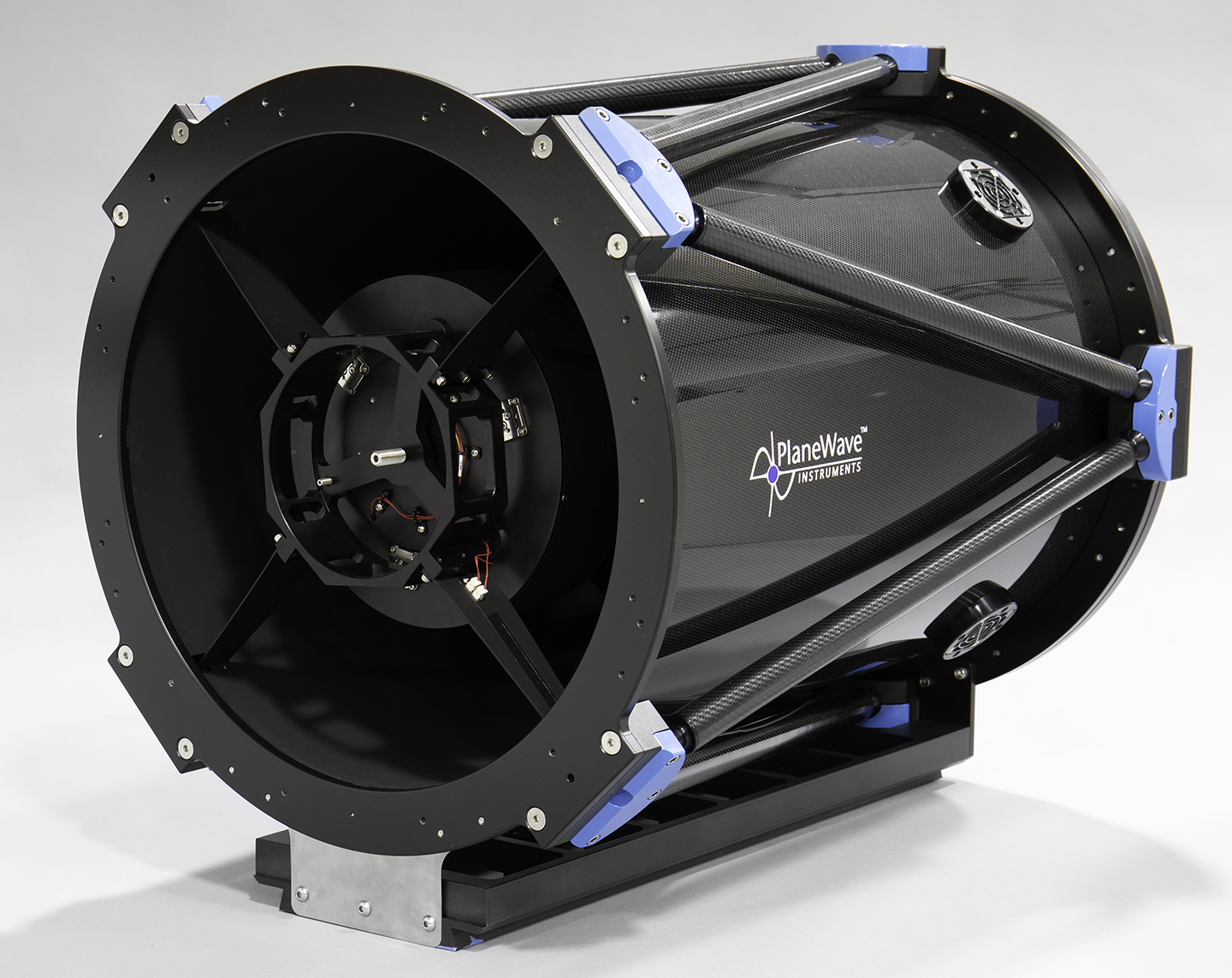}}
\subfigure[]{\includegraphics[width=0.36\textwidth]{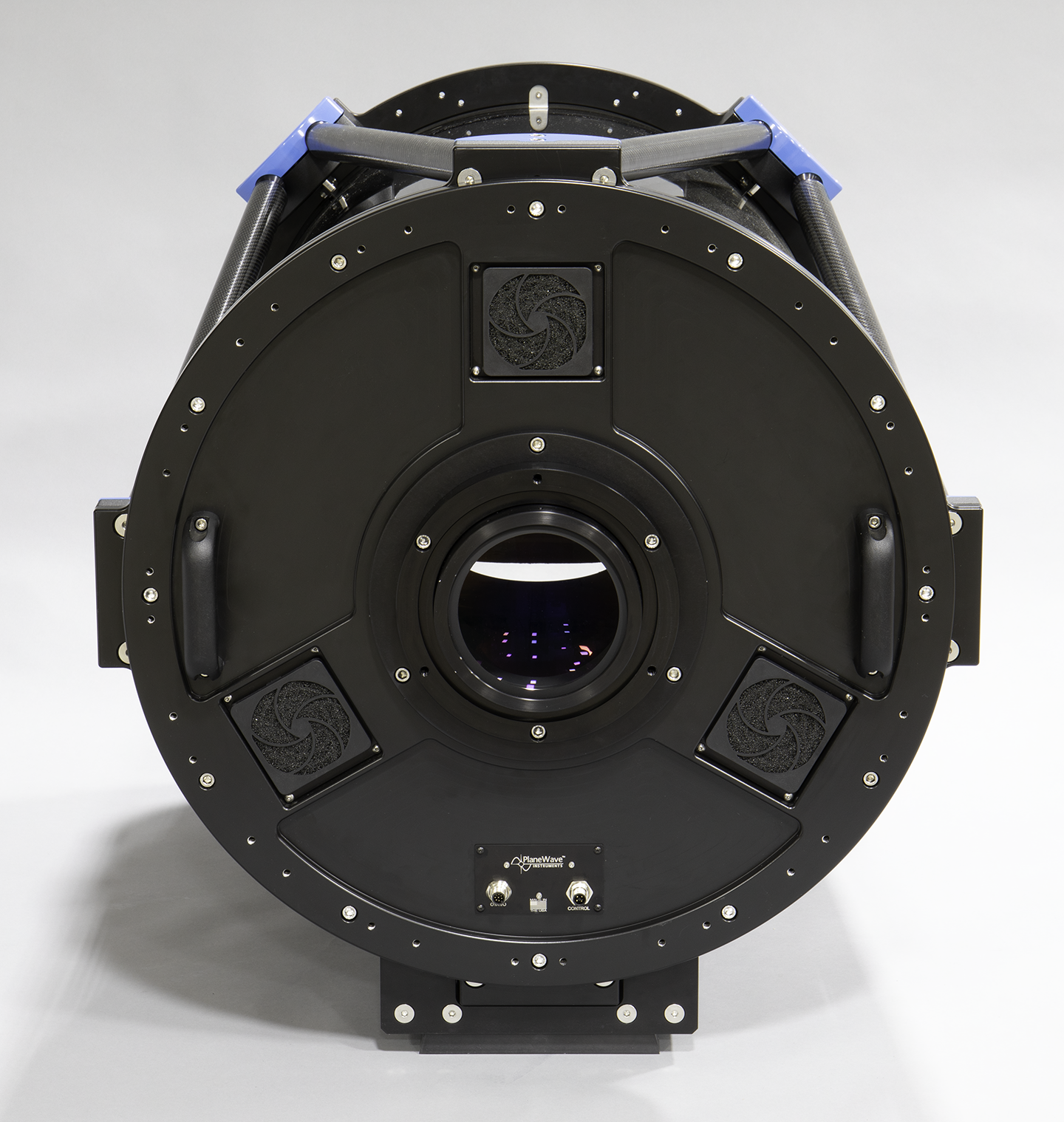}}

\end{tabular}
\end{center}

\caption[example] 
{\label{fig:dr500} 
Pictures of DeltaRho 500 (DR500). (a) A view from the front side showing the secondary mirror and its spider structure.
(b) A view from the backside.}

\end{figure} 
\subsection{Optical Tube Assmebly}
7DT consists of 20 units of 0.5-m telescope named DeltaRho 500 (DR500) which is manufactured by the PlaneWave Instruments.
Adopting a corrected Cassegrain optical design, DR500 has an elliptical primary mirror with a diameter of 508 mm and a spherical secondary mirror with a diameter of 286 mm.
In front of its Cassegrain focus, it employs a lens group consisting of 3 lenses that correct field curvature providing an image circle of 70 mm.
With its focal length of 1537 mm, its focal ratio is f/3 and its image circle of 70 mm covers about 2.6$^\circ$.
More information on DR500 is available at PlaneWave Instruments' website\footnote{\url{https://planewave.com/product/dr500-observatory-system/}}.

\subsection{Mount}

Each DR500 unit is mounted on an L-500 mount which is also manufactured by PlaneWave Instruments.
Employing direct-drive motors on its axes, the L-500 can function as an Alt/Az configuration as well as an equatorial configuration using an edge.
DeltraRho 500 units of 7DT use equatorial wedges and thus function as equatorial systems.
More information on L-500 is also available at PlaneWave Instruments' website\footnote{\url{https://planewave.com/product/l-500-direct-drive-mount/}}.

\subsection{Camera}

We utilze C3-61000 PRO cameras which are manufactured by Moravian Instruments.
This CMOS camera employs a back-illuminated SONY IMX455 sensor which has 9576 by 6388 pixels of 3.76 $\mu m$.
At the focal plane of DR500, the sensor covers 1.34$^\circ$ by 0.90$^\circ$.
The camera utilizes a rolling shutter function and its base readout noise is 3.51e- RMS.
More information on C3-61000 PRO camera is available at Moravian Instruments' website\footnote{\url{https://www.gxccd.com/art?id=647&lang=409}}.

\begin{figure} [ht]
\begin{center}
\begin{tabular}{c} 
\includegraphics[width=1.0\textwidth]{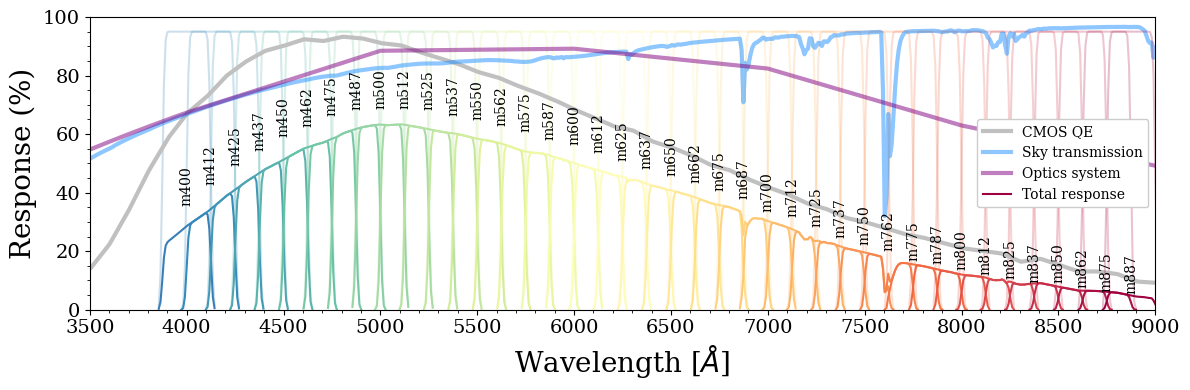}
\end{tabular}
\end{center}

\caption 
{\label{fig:filter} 
The filter response curves for 40 medium-band filters of 7DT.
These medium-band filters have an FWHM of 25nm and their central wavelengths have gaps of 12.5 nm.
They are designated by their central wavelengths.
For example, the filter covering the shortest wavelength range is m400 and its central wavelength is 400 nm, while the one covering the longest wavelength range is m887 and its central wavelength is 887.5 nm.
These response curves take into account the detector's quantum efficiency, sky transmission, telescope optics, and filter transmission.}

\end{figure} 

\subsection{Filter}
Each camera has a filter wheel that houses 9 filters, 3 of which are occupied by Sloan g-, r-, and i-band filters.
There are additional Sloan filters available for several units of DR500; 1 u-band filter and 3 z-band filters, respectively.
These Sloan filters are also commercial filters manufactured by Chroma Technology Corporation.

On top of these broad-band filters, 7DT has 40 medium-band filters which have an FWHM of 25 nm.
The central wavelengths of these medium-band filters range from 400 nm to 887.5 nm and the gaps between them are 12.5 nm.
We designate these medium-band filters by their central wavelengths; for example, m400 is the shortest medium-band filter whose central wavelength is 400 nm.
At the moment, there are 20 medium-band filters in use.
Instead of 12.5 nm, these 20 medium-band filters have gaps of 25 nm between them; thus m400 covers the shortest wavelength, while m875 covers the longest wavelength.
These medium-band filters are manufactured by Edmund Optics.

\subsection{Site}

7DT is hosted at El Sauce Observatory in Rio Hurtado Valley which is located at 30$^\circ$28$'$16$''$ S and 70$^\circ$45$'$47$''$ W.
The site of 7DT is close to those of Cerro Tololo Inter-American Observatory (CTIO), Gemini South Telescope, the Southern Astrophysical Research Telescope, and the Vera C. Rubin Observatory and shares excellent sky conditions with them.
The typical seeing of the site is about 1.5$''$ and the annual average number of clear nights is 300.
The mean sky brightness at the zenith measured by a sky quality meter is 21.97 mag arcsec$^{-2}$.
The infrastructure and maintenance of the site are provided by a Chilean telescope hosting company, ObsTech.

\begin{figure} [ht]
\begin{center}
\begin{tabular}{cc} 
\subfigure[]{\includegraphics[width=0.5\textwidth]{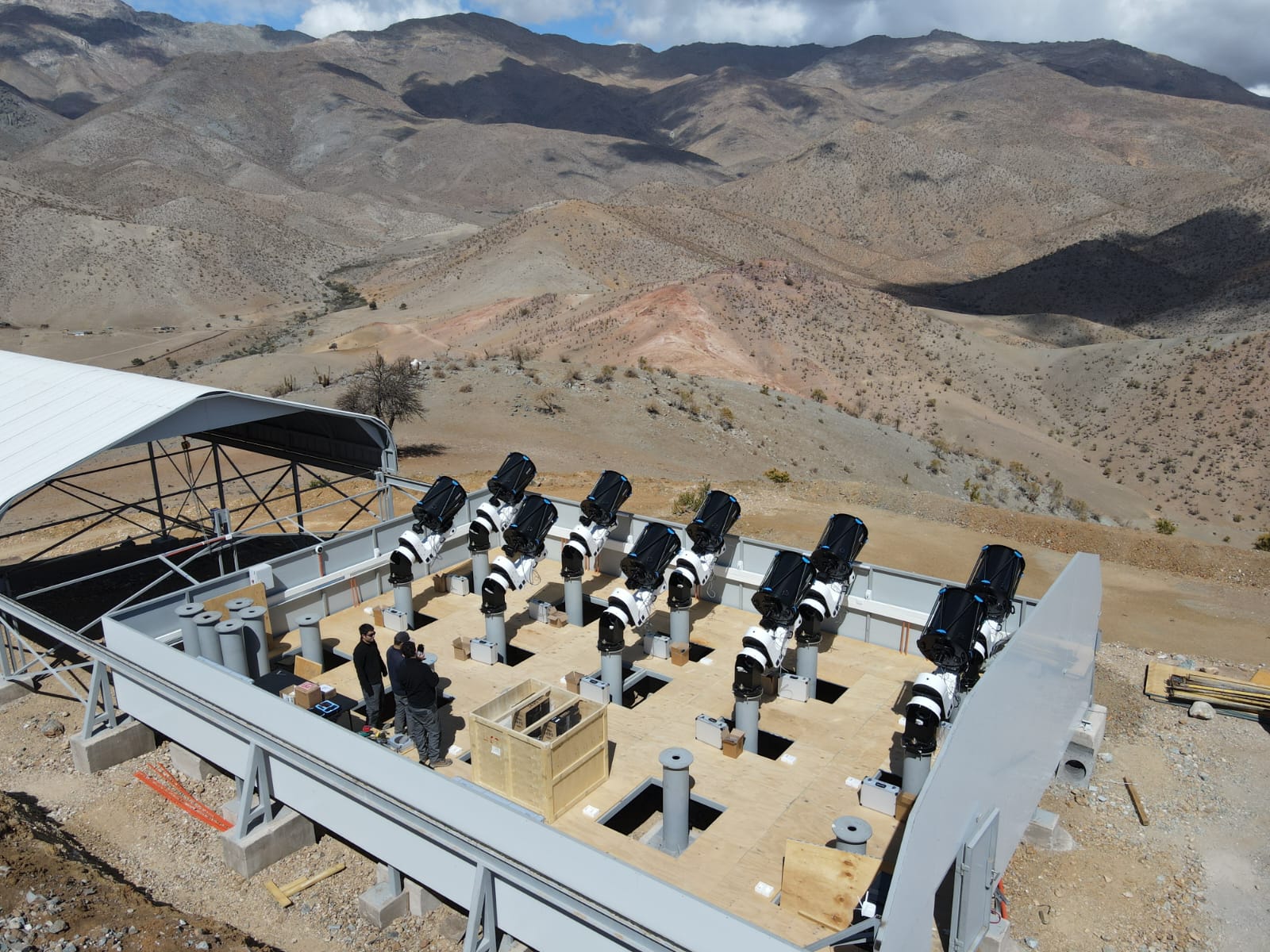}}
\subfigure[]{\includegraphics[width=0.48\textwidth]{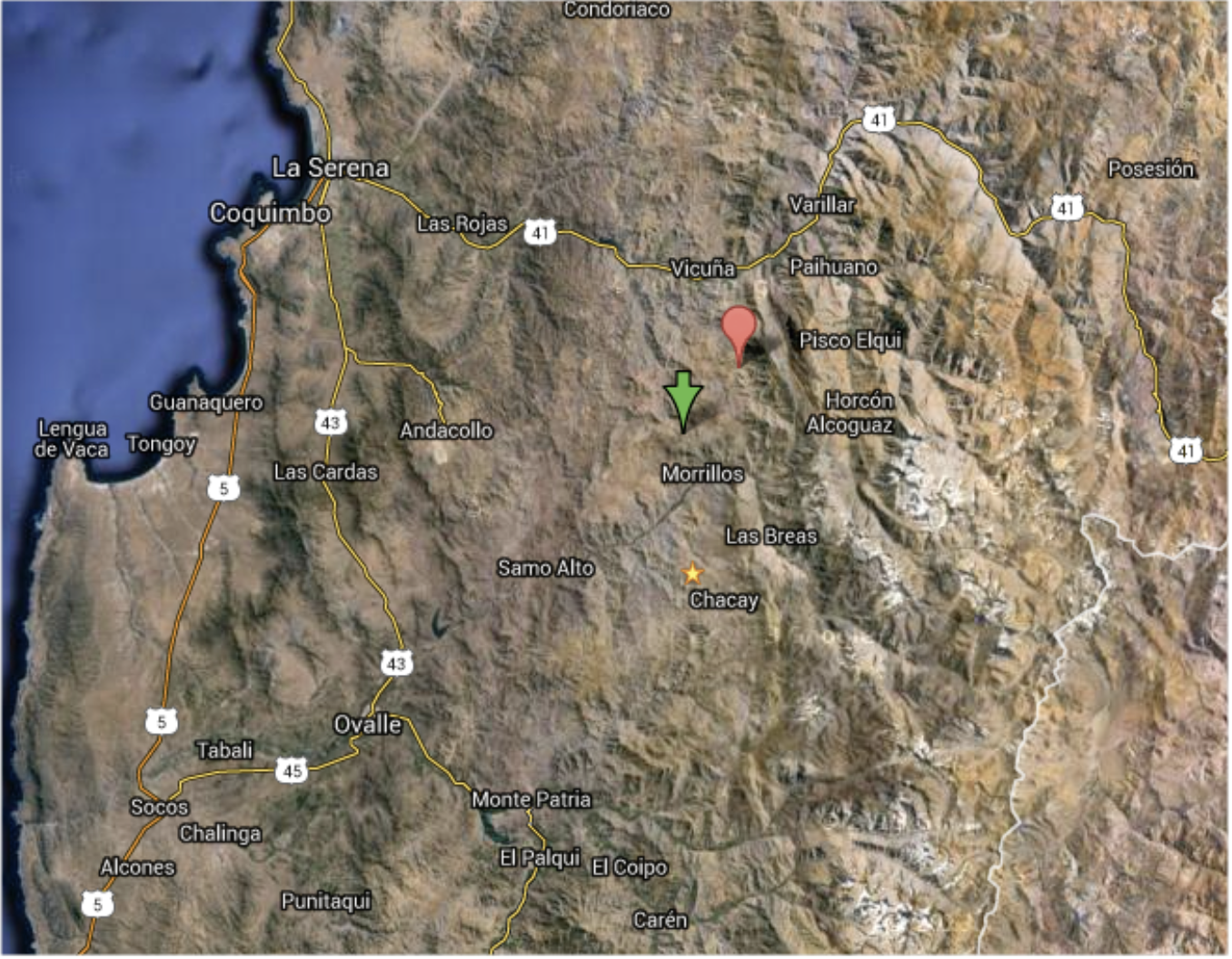}}

\end{tabular}
\end{center}

\caption 
{\label{fig:site} 
(a) A bird's eye view of 7DT at El Sauce Observatory in Chile. The picture shows 10 units of DR500 installed as of September 2023.
(b) The location of the El Sauce Observatory is indicated with a yellow star. Green and Orange arrows show the respective locations of CTIO and Cerro Pachon.}

\end{figure} 

\begin{table}[ht]
\caption{7DT System Specifications} 
\label{tab:spec}
\begin{center}       
\begin{tabular}{lc}
\hline
\hline
\rule[-1.5ex]{0pt}{4.0ex}  Property & Value \\
\hline
\hline
\multicolumn{2}{c}{\rule[-1ex]{0pt}{4.0ex} \bf{Unit Telescope}}\\
\rule[-1ex]{0pt}{3.5ex}  OTA design & Corrected Cassegrain \\
\rule[-1ex]{0pt}{3.5ex}  Primary diameter & 50.8 cm \\
\rule[-1ex]{0pt}{3.5ex}  Primary shape &  Prolate Ellipsoid\\
\rule[-1ex]{0pt}{3.5ex}  Secondary diameter &  28.6 cm\\
\rule[-1ex]{0pt}{3.5ex}  Secondary shape &  Spherical\\
\rule[-1ex]{0pt}{3.5ex}  Focal length &  153.7 cm\\
\rule[-1ex]{0pt}{3.5ex}  Focal ratio &  f/3.0\\
\rule[-1ex]{0pt}{3.5ex}  Image circle &  70 mm\\
\multicolumn{2}{c}{\rule[-1ex]{0pt}{4.0ex}\bf{Mount}}\\
\rule[-1ex]{0pt}{3.5ex}  Mount design & Alt-Azimuth/Equatorial Direct Drive \\
\rule[-1ex]{0pt}{3.5ex}  Mount slew rate & 20 degrees per second \\
\multicolumn{2}{c}{\rule[-1ex]{0pt}{4.0ex}\bf{Camera \& Filter}}\\
\rule[-1ex]{0pt}{3.5ex}  Sensor &  IMX455\\
\rule[-1ex]{0pt}{3.5ex}  Sensor size &  36 $\times$ 24 mm \\
\rule[-1ex]{0pt}{3.5ex}  Sensor dimension &  9576 $\times$ 6388 \\
\rule[-1ex]{0pt}{3.5ex}  Pixel size &  3.76 $\mu m$ \\
\rule[-1ex]{0pt}{3.5ex}  Pixel scale &  0.5$''$ \\
\rule[-1ex]{0pt}{3.5ex}  Field of view & 1.34 degree $\times$ 0.90 degree \\
\rule[-1ex]{0pt}{3.5ex}  Gain &  0.18-0.80 e$^{-}$/ADU \\
\rule[-1ex]{0pt}{3.5ex}  Readout noise &  1.39-3.51 e$^{-}$ \\
\rule[-1ex]{0pt}{3.5ex}  Full-well capacity &  11600-52800 e$^{-}$ \\
\rule[-1ex]{0pt}{3.5ex}  Non-linearity &  $<$0.2 per cent \\
\rule[-1ex]{0pt}{3.5ex}  Broad-band filter & Sloan u, g, r, i, z \\
\rule[-1ex]{0pt}{3.5ex}  Medium-band filter & m400, m413, m450, ..., m875 \\
\multicolumn{2}{c}{\rule[-1ex]{0pt}{4.0ex}\bf{Site}}\\
\rule[-1ex]{0pt}{3.5ex}  Lattitude & 30$^\circ$28$'$16$''$ S  \\
\rule[-1ex]{0pt}{3.5ex}  Longitude & 70$^\circ$45$'$47$''$ W \\
\rule[-1ex]{0pt}{3.5ex}  Altitude &  1600 m \\

\hline
\end{tabular}
\end{center}
\end{table}

\subsection{Computing Facility}

The 7DT computing facility is designed to provide a data inflow capacity of up to 1 TB per night for our planned survey described in Sec.~\ref{ss:survey} and to support the detection of fast transients as close to real-time as possible.
We set up a data storage system with a volume of 1PB and plan to increment the volume of our data storage by 1PB per year to at least 5PB.
A single raw image taken by each telescope unit is ~117 MB in size and is first transferred from an individual computer allotted to each telescope unit, Telescope Control Computer (TCC) to a server unit,  Main Control Computer (MCC) located at El Sauce Observatory. 
The data transfer from MCC is then handled through a Chilean private network provider, which is connected to a high-speed network connection to the Korea Research Environment Open NETwork (KREONET)\footnote{\url{https://www.kreonet.net}}.
Operated and maintained by Korea Institute of Science and Technology Information (KISTI)\footnote{\url{https://www.kisti.re.kr/eng/}}, KREONET provides high-performance network infrastructure and connectivity.
Furthermore, we implemented a high-performance data processing system with two GPU cards (NVIDIA A100) for real-time data reduction and analysis. 

\section{Software}
\label{sec:software}

\subsection{Telescope Control System}

The requirements for the operation system of 7DT as GW event follow-up facility and survey facility demand a lot.
It must support diverse observation modes of 7DT, and carry out synchronized operation of multiple units of telescopes.
It also must promptly respond to requests for target of opportunity (ToO) observations.

7DT has three operation modes; the Spec mode, the Deep mode, and the Search mode.
The Spec mode, abbreviated from the spectroscopic observation mode, points the entire telescope array to an identical sky position with different medium-band filters.
The Deep mode also points the telescope array to a sky position but with a single filter, presumably a broad-band filter.
In the Search mode, each unit points to a different sky position covering as large a sky area as possible with a broad-band filter.

We develop a software suite, Telescope Control System with Python (TCSpy), to implement these capabilities.
TCSpy is a Python-based software package designed for the synchronized control of ASCOM(Astronomy Common Object Model)-based multi-telescope array.

TCSpy utilizes ASCOM Alpaca\footnote{\url{https://www.ascom-alpaca.org}} and PlaneWave Interface 4 (PWI4)\footnote{\url{https://planewave.com/software/}} HTTP API to ensure stable communication within the telescope array based on HTTP protocol.
TCSpy comprises three types of modules; Device module, Action module, and Utility module.
Device modules allow TCSpy to connect with various components of the telescope array system.
Action modules provide the operation practices, organized into four levels based on the complexity of the operations.
Utility modules provide essential support, such as image header control, data I/O, and target database control. 
Combining these modules, TCSpy generates Applications which are defined as lists of actions for automated operations.

TCSpy incorporates an interactive target database based on MySQL database and a scheduling system.
The Target database stores targets, updates their observing status, and evaluates target rankings based on a scoring algorithm.
The scoring algorithm takes into consideration altitudes, moon separations, and observing priorities.
When ToO alerts are received, either from brokers or authorized personnel, the Target database puts ToO as the highest priority during the automatic target update and selects ToO as the optimal target.

The more technical and practical details of TCSpy are presented by Choi et al. (13101-106) in this Proceedings series.

\subsection{Data Reduction Pipeline}

The data reduction is performed by a modified version of \texttt{gpPy}.
The original \texttt{gpPy} (\url{https://github.com/SilverRon/gppy})\cite{2023zndo...8318777P}  is an automatic data reduction package developed to handle optical and NIR images from various observatories for transient searches\cite{2024ApJ...960..113P}.
It performs basic reduction, astrometric calibration, multiple image stacking, photometric calibration, and transient searches using Differential Image Analysis (DIA).
Thanks to the large number of pixels, 7DT can cover a wide field of view with fine pixel resolution. However, the size of a single frame is more than 100 MB per image, which significantly increases the computing time, especially in array calculations.

The modified \texttt{gpPy} (\texttt{gpPy+GPU}) is designed to utilize a GPU-accelerated library for array calculations and is specialized to process 7DT data. The basic procedures of the \texttt{gpPy+GPU} are as follows:

\begin{enumerate}

\item \textbf{Data Upload Monitoring}: Monitor newly uploaded data.
\item \textbf{Master Frame Creation}: Generate and save Master Bias, Master Dark, and Master Flat frames using the GPU-accelerated library.
\item \textbf{Data Reduction}: Perform Bias, Dark, and Flat corrections using the GPU-accelerated library.
\item \textbf{Astrometry}: Use \texttt{astrometry.net}\footnote{\url{https://astrometry.net}} and \texttt{SCAMP} \cite{2006ASPC..351..112B} to determine the celestial coordinates.
\item \textbf{Photometry}: Run photometry on reduced single frames using \texttt{SExtractor} \cite{1996A&AS..117..393B}.
\item \textbf{Image Alignment and Combination}: Align and combine images using \texttt{SWarp}\cite{2002ASPC..281..228B}.
\item \textbf{Zeropoint Calibration}: Calibrate zero points using point sources in the image with synthetic photometries from Gaia XP spectra (detailed in section \ref{ss:SPCalib}).
\item \textbf{Image Subtraction}: Use \texttt{HOTPANTS} \cite{2015ascl.soft04004B} for image subtraction.
\item \textbf{Transient Search}: Filter transient candidates and generate stamp images.
\item \textbf{Real/Bogus Classification}: Classify candidates with a CNN-based machine learning model.

\end{enumerate}

The GPU application significantly reduces the computing time for data reduction, making it more than ten times faster than the original \texttt{gpPy}.
Given that data from 10 units (during the commissioning stage) is uploaded simultaneously, we set the batch size to account for limited GPU memory.
To flexibly operate and maximize the capabilities of the 20 independent units, \texttt{TCSpy} includes various observing modes, such as deep and wide search modes.
Therefore, functions to handle these modes automatically will be added to \texttt{gpPy+GPU}.

\section{Commissioning Procedure}

The commissioning procedures of observational instruments consist of twofold; evaluating the functionality of instruments and verifying their science capability.
To test the functionality of instruments, optomechanical alignment and polar alignment were evaluated as well as instrument foci and quality of image quality were assessed.
To verify scientific capability, we performed various observations to obtain data for basic calibration, spectrophotometric calibration, and pilot studies for science cases.

\subsection{Instrument Functionality Evaluation}
To ensure its performance, each unit of DR500 underwent several optical alignment procedures; collimating its optical path and adjusting its focal plane.
The focal plane alignment requires great attention due to its optomechanical structure; a camera is attached to a tip/tilt adaptor which is adjusted by three screws.
Even small slight misalignment between focal planes of the optical instrument and the camera affects overall image quality across the image field.
We apply a couple of methods to align them; comparing focus test images, and evaluating astigmatisms and aberration across the field.

Polar alignment is carried out by the pointing modeling procedure of PlaneWave Interface 4 (PWI4), the main control software for mounts produced by PlaneWave Instruments.
In general, 40 to 50 sky points between 60 and 88 degrees latitude are taken for a pointing model.
While we do not regularly update pointing models, we monitor each unit's pointing and tracking accuracies and update one's pointing model whenever necessary. 

\subsection{Calibration Data Acquisition}

We take bias, dark, and flat frames as often and regularly as possible.
During the early phase of the commissioning, we took a huge number of bias frames to characterize each camera while monitoring the temperatures of the cameras.
Now we decided to take 18 bias frames each 30 minutes after the cooling procedure initiates and after the last science data is taken while the temperatures of the cameras are set to be -10$^\circ$, although bias levels are quite consistent with this temperature setting.
We found that bias levels are consistent with the manufacturer's specification being around 3.5 e$^{-}$ RMS.
We also confirmed that bias frames show the horizontal patterns which are known for CMOS detectors.
These horizontal patterns vary and do not show any dependency.

We also take 18 dark frames at the end of each night, although the levels of dark frames are quite low even for long ($>$100 seconds) exposures.

We take flat frames of every filter that we use for each night.
We took and evaluated flat frames to check the behaviors of combined flats for each camera and filter during the early phase of the commissioning.
At this point, it is unclear if combined flats over long periods can appropriately perform the flat fielding.
Therefore, we decided to take nine flat frames at the end of night operations.

\subsection{Spectrophotometric Calibration}
\label{ss:SPCalib}

Gaia XP spectra from the all-sky \cite{2023A&A...674A...1G} provide homogeneous calibration for optical surveys \cite{2023ApJS..268...53X, 2024A&A...683A..29L}, but dependencies on both color and brightness have been reported by Ref.~\citenum{2023A&A...674A..33G}.
Therefore, we corrected these dependencies as detailed in Ref.~\citenum{2024A&A...683A..29L}.
The homogenization procedures consist of three steps:
First, we obtained medium-band images of 68 spectrophotometric standard stars (SPSS), including CALSPEC sources.
For these frequently used standard stars, we requested synthetic photometry based on the filter response curves of 7DT's medium-band filters and Gaia XP spectra from Gaia DR3\cite{2021A&A...649A...1G}.
Invariable point sources were filtered from the provided Gaia catalogs following the criteria of the SkyMapper Survey DR4 \cite{2024arXiv240202015O}.
Finally, we calibrated the obtained images and corrected the magnitude offset dependency between the calibrated magnitude and synthetic photometry from Gaia XP on the Gaia \(G\)-band magnitude and the color \(G_{BP} - G_{RP}\) iteratively.
As a result, error budgets range approximately from 4 to 12 mmag.
We use differential photometry to define the zero-point of the image.
The spectrophotometric calibration procedures consist of three steps.
We calibrated the instrument magnitude against the corrected synthetic photometry of matched Gaia sources using a 3-sigma clipping of zero points.
These processes are repeated for various aperture sizes. The details of both the correction and calibration procedures will be provided by Paek et al. ({\it in prep}).

\subsection{Science Verification}
Although the main scientific goals of 7DT involve GW events and their electromagnetic counterparts, 7DT will provide low-resolution spectroscopic datasets that provide opportunities to explore numerous scientific objectives.
The list of these goals includes, but is not limited to time-domain studies of the Solar System object, young stellar objects and active galactic nuclei, resolved stellar populations of nearby galaxies, high-redshift galaxy clusters, and large-scale structure.
Since October 2023 when the first light image of 7DT was taken, we have been carrying out several observational projects to verify the science capability of 7DT.
We list some of the projects to enlighten prospects of the scientific capabilities of 7DT.

We obtained the medium-band images of two main-belt asteroids, 13 Egeria and 10 Hygiea, to detect the 0.7 $\mu  m$ absorption feature.
The presence of the 0.7 $\mu m$ absorption feature reveals the Ch-type asteroids, which is important to trace the thermal history of the hydrated asteroids.

We observed the field of Orion Molecular Cloud 2 and 3 (OMC2/3) region at two epochs with a 1-day cadence, while we also observed the Serpens Main Molecular Cloud at three epochs.
The data of OMC2/3 yields three new T Tauri stars two of which show H$_{\alpha}$ emission feature, which shows that multi-epoch monitoring of star-forming regions by 7DT is capable of discovering and characterizing the early phase of stellar evolution.

Farther out, we observed 40 galaxies drawn from the sample of the PHANGS (Physics at High Angular Resolution in Nearby Galaxies) survey. 
A multi-wavelength survey effort, the PHANGS survey consists of several surveys carried out by the Atacama Large Millimeter/submillimeter Array, JWST, the Hubble Space Telescope,  the Very Large Telescope (VLT), and AstroSat covering the wavelength range from mm to far ultraviolet.
Among these, PHANGS-MUSE based on IFU (integral field unit) data from MUSE (the Multi Unit Spectroscopic Explorer) of VLTprovides good references to compare and evaluate the 7DT data.

Estimating more precise photometric redshifts is one of the advantages of 7DT thanks to the better spectral resolution compared to ones using broad-band photometry.
We anticipate assembling a photometric catalog that yields highly reliable photometric redshift estimation.
In this regard, we carried out two observations; one for a galaxy cluster, PSZ G227.44-31.24, and the other for the COSMOS field.
Photometric redshifts for galaxies within the fields will be estimated and compared to known spectroscopic redshifts up to z$\sim$0.5.

Last but not least, we had opportunities to test the capabilities of 7DT as a transient follow-up facility.
We searched for the optical counterpart of the bright X-ray flare, EPS20240219aa\cite{2024GCN.35773....1Z} using 7DT and reported the upper limits for several medium-band filters\cite{2024GCN.35790....1P}.
We also conducted a follow-up observation for the LVK GW event, S240422ed\cite{2024GCN.36263....1K} covering $\sim$170 square degrees with Sloan r-band and the suite of medium-band filters\cite{2024GCN.36352....1C}.


\subsection{Current Performance}
We have been assessing the optical performance of 7DT since we commenced the commissioning procedures in October 2023.
After collimation and tip-tilt correction, all the units of DR500 show reasonable image qualities based on the measured FWHM across the FoV.
Most units show that they can achieve better than 1.5$''$ on good nights.
Considering the fact that the median seeing at the site is 1.5$''$, these are reasonable performance results.

The typical 5$\sigma$ limiting magnitudes with the fiducial exposure time of 120 seconds are about 20.2 mag and 19.1 mag for Sloan r-, and i-bands, respectively.
For stacked images with 1800 seconds, the limiting magnitudes are about 21.1 mag, 21.0 mag, and 20.1 mag for Sloan g-, r-, and i-bands, respectively.
For selected medium-band filters, the limiting magnitudes are 19.1 mag, 18.7 mag, and 17.4 mag for m600, m700, and m800 filters, respectively.
These image depths are still preliminary measurements.

\section{Prospects}

\subsection{Future Schedule}
As of May 2024, 12 units of DeltraRho 500 are deployed at the site, El Sauce Observatory, and are operational.
Four more units are scheduled to be deployed by the end of June, or early July.
These additional units will undergo extensive commissioning procedures as the units installed already.
CMOS cameras for these units are already ready.

We are also exploring options to procure additional medium-band filters whose central wavelengths fill in the gaps between the current medium-band filter set.

\subsection{Survey Plan}
\label{ss:survey}

We will soon commence regular surveys.
Our survey plan consists of three layers; Reference Imaging Survey (RIS), Wide-Field Survey (WFS), and Intensive Monitoring Survey (IMS).

RIS will cover the majority of the Southern sky except the Galactic Plane.
Using the canonical exposure time of 100 seconds, we will obtain three exposures per visit, which makes a total exposure per visit 5 minutes.
For RIS, we plan to visit once for each sky tile generated by HEALpix for the first year of the survey period.
The total observational time budget for RIS is 50,000 minutes to cover about 20,000 square degrees.

WFS will visit each sky position within the survey area with a 14-day cadence during the survey period of five years.
The survey field of WFS is still being considered, although it is most likely to include fields with ancillary data, especially near-infrared photometry.
WFS will cover up to 1600 square degrees.

IMS will visit the survey field every day for the survey period.
We chose AKARI Deep Field South, ADF-S as the target field for IMS.
The observational time budget for IMS is 20000 minutes per year.

\acknowledgments 

The 7-Dimensional Telescope is designed, built, and operated by Center for the Gravitational-wave Universe at Seoul National University which is supported by the National Research Foundation of Korea (NRF) grant, No. 2021M3F7A1084525 by the Korea government (MSIT).
We also acknowledge the support from the National Research Foundation of Korea (NRF) grant, No. 2020R1A2C3011091.
This work was supported by KREONET(Korea Research Environment Open NETwork)/KISTI(Korea Institute of Science and Technology Information).
J.H.K acknowledges the support from the Institute of Information \& Communications Technology Planning \& Evaluation (IITP) grant, No. RS-2021-II212068.
SWC acknowledges the support from the National Research Foundation of Korea (NRF) grants, No. 2020R1A2C3011091 and No. 2021M3F7A1084525 funded by the Ministry of Science and ICT (MSIT) as well as Basic Science Research Program through the NRF funded by the Ministry of Education (RS-2023-00245013).

\bibliography{SPIE_13094} 
\bibliographystyle{spiebib} 

\end{document}